\documentclass[twocolumn]{aastex631}

\usepackage{color}
\usepackage{amsmath}
\definecolor{jaredpurple}{RGB}{93, 63, 211}

\newcommand{\Msun}{M_\odot}
\newcommand{\Rsun}{R_\odot}

\newcommand{\Menv}{M_{\rm env}}
\newcommand{\Mej}{M_\mathrm{ej}}

\newcommand{\Mten}{M_{10}}

\newcommand{\Menvten}{M_\mathrm{env,10}}
\newcommand{\Mejten}{M_\mathrm{ej,10}}

\newcommand{\Rfh}{R_{500}}

\newcommand{\Eexp}{E_\mathrm{exp}}

\newcommand{\Efoe}{E_{51}}
\newcommand{\Lfifty}{L_{50}}
\newcommand{\Lft}{L_{42}}
\newcommand{\tptwo}{t_{\rm p, 2}}
\newcommand{\Ni}{{^{56}{\rm Ni}}}

\newcommand{\MNi}{M_{\rm Ni}}
\newcommand{\tpt}{t_{\rm p}}
\newcommand{\days}{{\rm d}}

\shorttitle{Families for SN2023ixf}
\shortauthors{Forde \& Goldberg}
\graphicspath{{./}{figures/}}

\begin{document}

\title{Modeling Supernova 2023ixf: Lightcurve Degeneracies and Morphological Differences}

\correspondingauthor{Shannon Forde}
\email{shannon.forde97@bcmail.cuny.edu}

\author[0009-0005-5767-0034]{Shannon Forde}
\affiliation{CUNY Brooklyn College, Department of Physics, Brooklyn, NY, USA}

\author[0000-0003-1012-3031]{Jared A. Goldberg}
\affiliation{Center for Computational Astrophysics, Flatiron Institute, New York, NY, USA}

\begin{abstract}

Supernova 2023ixf is a type II-P supernova that was observed in May 2023 in the spiral galaxy Messier 101. This was the closest supernova observed of the decade, making this an exciting discovery. Combining the observed brightness and duration with theoretical scaling relations, we model the lightcurve of this supernova in order to reveal the properties of the progenitor star at the time of explosion, including its mass, radius, and explosion energy. We simulate these explosions using the stellar evolution and radiation-hydrodynamics codes MESA+STELLA. We find that SN2023ixf is not easily explained with “normal” stellar evolution, and only models with a small mass of H-rich ejecta can fit the lightcurve. We also find that the late time properties of the lightcurve are better fit by a higher initial-mass ($M\gtrsim 20M_\odot$) star with substantial mass loss during its lifetime, as compared to models with lower initial mass ($M\lesssim 13M_\odot$) and less mass loss.

\end{abstract}


\section{Introduction} \label{sec:intro}

A Type II supernova (SN) occurs when a massive ($\ge 8M_{\odot}$), hydrogen-rich star dies in a core-collapse event. Type IIP supernovae (SNe-IIP) are core-collapse SNe that have a plateau in the lightcurve, and are the most common type of core-collapse supernova \citep{Smith2011}. The progenitors of these supernovae are known to be red supergiants, with masses greater than $8\Msun$ \citep{Woosley2002,Smartt_2009}.

SN2023ixf is a type IIP supernova that was discovered in May 2023 \citep{Itagaki_2023} in the nearby spiral galaxy M101, which is 6.71 Mpc from Earth \citep{Riess2016}. The close proximity made it possible to conduct extensive observation of the SN, making it one of the most well-observed type II supernovae. If we are able to determine the zero-age main-sequence (ZAMS) mass of the progenitor star, it could help us to better understand stellar evolution and the explosions of stars. However, there has been a wide range of estimates for $M_\mathrm{ZAMS}$. While some have estimated masses as low as 8 - 10$M_{\odot}$ \citep[e.g.][]{Pledger2023,Moriya2024, Bersten2024}, others have much larger estimated masses of 16 - 20$M_{\odot}$ \citep[e.g.][]{Qin2024,Hsu2024, Fang2025}. The wide variety of estimates for $M_\mathrm{ZAMS}$ comes from the different methods used, such as direct observations of the red supergiant (RSG) progenitor before the explosion \citep[e.g.][]{Pledger2023,Soraisam2023, Jencson_2023,
Kilpatrick2023, VanDyk2024, Qin2024,Xiang2024}, or from modeling the supernova lightcurve and other observables \citep{Hiramatsu2023,Bersten2024,Moriya2024,Hsu2024,Fang2025}. Direct measurements are difficult because of the large amounts of circumstellar dust that obscured the progentitor. This fits with the data seen in the lightcurve and spectrum from early times in the SN, which points to a high mass of circumstellar material  \citep[e.g.][]{Hiramatsu2023, 
Smith_2023, Bostroem2023, Singh_2024,
Zimmerman2024, Martinez2024}.
A lot of attention has been paid in the literature to the early-time emission from interaction with the dense circumstellar environment \citep[e.g.][]{Berger2023, Bostroem2023, Hiramatsu2023, Zhang2023, 
Martinez2024,Chandra2024, Shrestha2025}. In this paper, we focus our analysis on the later-time properties of the lightcurve plateau of SN2023ixf to extract information about the mass and radius of the progenitor and the explosion energy from the data.

\section{Methods} \label{sec:methods}

For an observed Type IIP Supernova with an given luminosity at day 50 $(\Lfifty)$, plateau duration $(\tpt)$, and $\Ni$ mass 
($\MNi\gtrsim0.03\Msun$), 
the mass of the H-rich ejecta $(\Mten\equiv\Mej/10\Msun)$ and explosion energy $(\Efoe\equiv\Eexp/10^{51}\mathrm{ergs})$ 
can be determined as a function of progenitor radius ($\Rfh\equiv R/500\Rsun$) 
via the following scaling relations \citep{Goldberg2019}:
\begin{equation}
\begin{split}
\begin{aligned}
\log(\Efoe)&=-0.728+2.148\log(\Lft)-0.280\log(\MNi) \\&+2.091\log(\tptwo)-1.632\log(\Rfh),\\
\log(\Mejten)&=-0.947+1.474\log(\Lft)-0.518\log(\MNi)
\\&+3.867\log(\tptwo)-1.120\log(\Rfh),
\label{eq:scaling}
\end{aligned}
\end{split}
\end{equation}
where $\MNi$ is in units of $\Msun$, $\Lft=\Lfifty/10^{42}$ erg s$^{-1}$, 
and $\tptwo=\tpt/100\,\days$. 

In order to apply these scalings to SN2023ixf, bolometric luminosity data was taken from \citet{Zimmerman2024}\footnote{using WebPlotDigitizer \citep{WebPlotDigitizer}}. 
From this data, the end of the bolometric lightcurve plateau was fit from day 50 until day 110 with the functional form 
\begin{equation}
y(t) = \frac{-a_\mathrm{0}}{1+e^{(t-t_\mathrm{p})/w_\mathrm{0}}}+(p_\mathrm{0})(t)+m_\mathrm{0} 
\end{equation}
following \citet{Valenti2016} using the procedure described in \citet{Goldberg2019} to find the plateau duration $\tpt$. Using a nickel mass $M_\mathrm{Ni}=0.075M_\odot$ following \citet{Zimmerman2024}, and $t_\mathrm{p}$ found from fitting the curve, the scaling equations (Eq.~\ref{eq:scaling}) recover the ejecta mass and the explosion energy of the star as a function of the stellar radius. We show the curves for $\Mej$ and the explosion energy as a function of the progenitor radius in the graphs in the left panels of Figure~\ref{fig:23ixf}.

A grid of stellar models constructed using 
\texttt{MESA} \citep{Paxton2011,Paxton2013,Paxton2015,Paxton2018,Paxton2019,Jermyn2023} from \citet{Goldberg2020}, with variations in the stellar wind efficiency, convective mixing length, initial rotation, core overshooting parameter, and initial mass, is overlaid on the graph of the ejecta mass, with darker colors indicating higher $M_\mathrm{ZAMS}$. As seen in the figure, there are very few stellar models from this grid that intersect with the scalings considering the ejecta mass of 2023ixf. 
In order to explode those models, we followed the test case \texttt{ccsn\_IIp} in \texttt{MESA} revision 24.03.1, where the explosion energy and $M_\mathrm{Ni}$ are free parameters. Chemical and density mixing during the shock propagation are treated with the \citet{Duffell_2016} prescription for mixing via the Rayleigh-Taylor Instability (RTI). After the shock reaches $xm=0.1M_\odot$ below the stellar surface in MESA, just prior to shock breakout, we evolve the ejecta further and model the lightcurves using the 1D radiation-hydrodynamics code \texttt{STELLA} \citep{Blinnikov1998, Blinnikov2004,Blinnikov2006,Baklanov2005}.\footnote{See \citet{Paxton2018, Goldberg2019,Goldberg2020} for further details on the numerical setup.} 

Because the hydrogen content of the envelope contributes to the plateau duration, a lower hydrogen-rich envelope mass may also be able to reproduce the brightness and duration of 2023ixf. We recalculated the scaling prefactor for the mass and the explosion energy to fit the envelope mass instead of the ejecta mass. To do this, we exploded six stripped envelope models ranging in birth mass from $19{\Msun}$ to $25{\Msun}$, in explosion energy from $0.36\times10^{51}$ erg to $0.88\times10^{51}$ erg, in envelope mass from $2.73\Msun$ to $5.13\Msun$, and in radii from $696\Rsun$ to $1167\Rsun$. These stellar models, including their envelope mass $M_\mathrm{env}=M_\mathrm{env}/10M_\odot$, $E_\mathrm{exp}$, $\Lfifty$, $\tpt$, $\MNi$, and $R$, were then used to calibrate the pre-factors for new scalings, following the same power laws as in Eq.~\ref{eq:scaling}:
\begin{equation}
\begin{split}
\begin{aligned}
\log(\Efoe)&=-0.341+2.148\log(\Lft)-0.280\log(\MNi)
\\&+2.091\log(\tptwo)-1.632\log(\Rfh),\\
\log(\Menvten)&=-0.836+1.474\log(\Lft)-0.518\log(\MNi)
\\&+3.867\log(\tptwo)-1.120\log(\Rfh).
\label{eq:scalingMenv}
\end{aligned}
\end{split}
\end{equation}
These scalings are shown in the middle left panel of Figure~\ref{fig:23ixf}, with the H-rich envelope masses from the same \citet{Goldberg2020} model grid overlaid \footnote{This progenitor model grid is publicly available at \href{https://github.com/aurimontem/MassiveStarGrid}{https://github.com/aurimontem/MassiveStarGrid}}. According to the envelope mass scalings, more stellar models could possibly fit the lightcurve of SN2023ixf.

\section{Results} \label{sec:methods}

\begin{figure*}
\includegraphics[width=\textwidth]{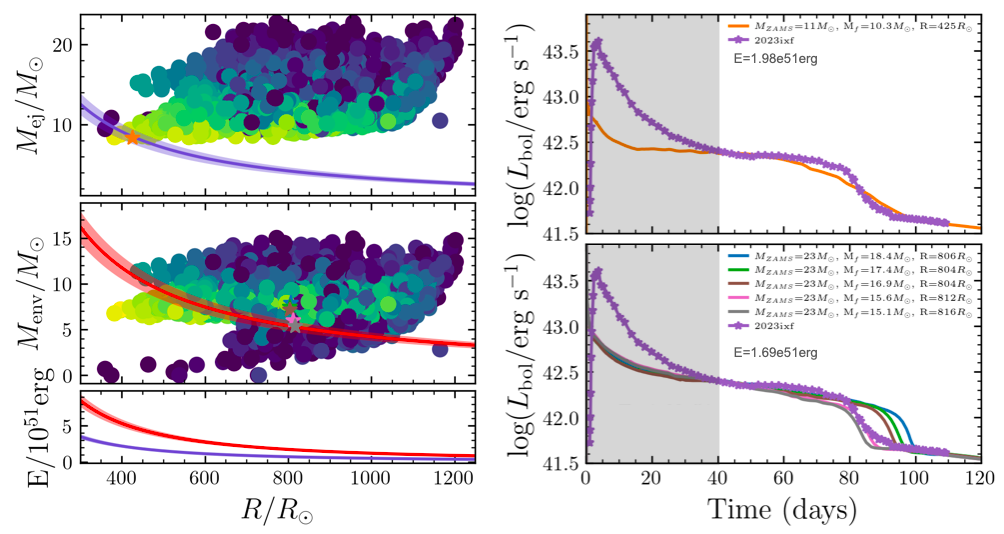}
\caption{Left Panels: Degeneracy curves for the ejecta mass (top), envelope mass (middle), and explosion energy (bottom) as a function of $R$ for SN2023ixf, with an 11 percent modeling error quoted in \citet{Goldberg2019} included as a shaded region around the curves. The model grid of stars obtained from \citet{Goldberg2020} is overlaid on the graphs of $M_{ej}$ and $M_{env}$, with color corresponding to $M_\mathrm{ZAMS}$ from 10-25$M_\odot$ (lighter colors indicate lower $M_\mathrm{ZAMS}$). Right Panels: The upper right panel shows the lightcurve for the explosion of an $M_\mathrm{ZAMS}=11M_{\odot}$ model taken from the $M_{ej}$ grid, compared to observations from \citet{Zimmerman2024} (purple starred lines). The lower right panel shows the lightcurves for $M_\mathrm{ZAMS}=23M_{\odot}$ models taken from the $M_{env}$ grid with different wind efficiency parameters. In the legend, $M_{ZAMS}$ is the birth mass of the star and $M_{f}$ is the mass at the time of the supernova. The grey shaded region indicates the circumstellar interaction phase, where we do not expect agreement between observations and the model lightcurves. 
\label{fig:23ixf}}
\end{figure*}

The right panels of Figure 1 show the bolometric lightcurves for SN2023ixf (purple stars) compared to synthetic lightcurves from \texttt{STELLA} for models selected and exploded from the different model grids (colored lines). The lightcurve indicates log of the luminosity as a function of time, where the time axis is the time since shock breakout. The upper right panel of Figure~\ref{fig:23ixf} shows the lightcurve from the explosion of a representative stellar model, indicated by the orange star in the upper left panel. This model was selected from our model grid consistent with the ejecta mass scaling (Equation 2) and the data from SN2023ixf. The model chosen from the grid had an $11{\Msun}$ initial mass, a $10.3{\Msun}$ final mass, a $425{\Rsun}$ radius, a helium core mass of $3.334{\Msun}$, a hydrogen-rich envelope mass of $\Menv=6.966{\Msun}$, 
an ejecta mass of $\Mej=8.35{\Msun}$,
and an explosion energy equal to $1.98\times10^{51}$erg. As seen in the figure, the plateau luminosity and duration from this model fit well with the overall brightness and duration of the SN2023ixf data, as well as the nickel tail of the lightcurve. However, at the end of the plateau, the shape of the model lightcurve does not match the SN2023ixf obeservations. The lightcurve from SN2023ixf has a steeper drop at the end of the plateau, while the stellar model has a more gradual decrease in brightness.\footnote{We reproduced this result with all stellar models from the grid which were consistent with Eq.~\ref{eq:scaling}, but show just one in Fig.~\ref{fig:23ixf} for simplicity.
The high-mass models denoted by the dark-blue points which intersect the $\Mej$-calibrated scalings showed IIb-like lightcurves due to their extremely low Hydrogen masses.}

The lower right panel of Figure~\ref{fig:23ixf} shows the lightcurves of a few higher mass stellar models taken from the grid based on the envelope mass scalings (Eq.~\ref{eq:scalingMenv}). All of the models shown have the same birth mass, $M_\mathrm{ZAMS}=23{\Msun}$, and explosion energy, $1.69\times10^{51}$ erg, but have varying winds. Although these stellar models have a higher birth mass than the $11{\Msun}$ model, they do have lower envelope masses. The blue line shows a model that has a final mass of $18.4{\Msun}$, a radius of $806{\Rsun}$, a hydrogen-rich envelope mass of ${8.68\Msun}$, and an ejecta mass of ${14.4\Msun}$. The green line shows a model with a final mass of $17.4{\Msun}$, a radius of $804{\Rsun}$, an envelope mass of ${7.76\Msun}$, and an ejecta mass of ${15.5\Msun}$. The brown line shows a model with a final mass of $16.9{\Msun}$, a radius of $804{\Rsun}$, an envelope mass of ${7.38\Msun}$, and an ejecta mass of  
${15.1\Msun}$. The pink line shows a model with a final mass of $15.6{\Msun}$, a radius of $812{\Rsun}$, an envelope mass of ${5.97\Msun}$, and an ejecta mass of  ${13.5\Msun}$. The gray line shows a model with a final mass of $15.1{\Msun}$, a radius of $816{\Rsun}$, an envelope mass of ${5.48\Msun}$, and an ejecta mass of ${13.2\Msun}$. As seen in the lower right panel of Fig.~\ref{fig:23ixf} comparing these model lightcurves to 2023ixf observations, the shape at the end of the plateau fits that of SN2023ixf much better than the $11{\Msun}$ model, specifically at the end of the plateau, where the steep drop is naturally reproduced by our calculations. Of the models shown, those with more wind, and therefore lower mass at the time of explosions, fit the data from 2023ixf better than the models with less wind. This trend continues up to a point, evident in the lower right panel where the gray line ($M_\mathrm{env}=5.48M_\odot$), which has the lowest mass at the time of explosion in this data set, does not fit the lightcurve of SN2023ixf as well as the pink line ($M_\mathrm{env}=5.97M_\odot$) does.

\section{Discussion} \label{sec:discussion}

We can conclude that although a low-mass star undergoing normal stellar evolution can match the luminosity of 2023ixf at day 50, as well as the duration of the plateau relatively well, the plateau lightcurves of 2023ixf are better modeled by a more massive star with higher mass loss during its lifetime and a higher core-to-envelope mass ratio. This is the first work to find both low-mass, unstripped stellar models and higher-mass, partially-stripped-stellar models which broadly match the observations from the same grid of progenitor models. This is guided by the comparison of the models to both sets of scaling relations considering the total ejecta mass and pre-explosion H-rich envelope mass independently, and the fact that the progenitor model grid is sufficiently extensive to include stellar models with a range of degrees of envelope stripping and a variety of masses and radii. These results are in agreement with more recent results that combine lightcurve modeling with other observables, including progenitor variability \citep{Hsu2024}, late-time nebular spectra \citep{Fang2025}, and are consistent even with detailed modeling of the combined early-phase and late-phase lightcurve \citep{Kozyreva2024}, which also requires a larger degree of envelope stripping. In the era of the Vera C. Rubin Observatory Legacy Survey of Space and Time (LSST), there will be many lightcurves that lack additional constraints, and this type of modeling could reveal and disentangle the degeneracies for many different supernovae.

\begin{acknowledgments}
We are grateful for many valuable conversations with members of the SSTARS group at CCA: Lieke van Son, Floor Broekgaarden, Isabelle Jones, Melanie Santiago, Ana Lam, and Soumendra Roy. We also acknowledge valuable correspondence with Brian Hsu and Daichi Hiramatsu about the observations. We likewise thank Tim Paglione. 
S.F. acknowledges AstroCom NYC, which is supported by the Simons Foundation and the National Science Foundation.
J.A.G. is supported by a Flatiron Research Fellowship. The Flatiron Institute is supported by the Simons Foundation. 
\end{acknowledgments}

\software{
Modules for Experiments in Stellar Astrophysics \citep[\texttt{MESA};][]{Paxton2011,Paxton2013,Paxton2015,Paxton2018,Paxton2019,Jermyn2023}, 
\texttt{STELLA} \citep{Blinnikov1998,Blinnikov2004,Baklanov2005,Blinnikov2006}, \texttt{numpy} \citep{Numpy}, \texttt{scipy} \citep{Scipy}, and \texttt{matplotlib} \citep{Matplotlib}.}

\bibliographystyle{aasjournal}
\bibliography{main.bib,HsuReferences.bib}

\end{document}